\documentclass[11pt,a4paper]{article}
\usepackage{amsfonts,amssymb}
\usepackage{cite}
\usepackage[T2A]{fontenc}
\usepackage[cp1251]{inputenc}
\usepackage[english]{babel}
\usepackage{amsmath}%
\usepackage[unicode]{hyperref}

\title{\bf On the recursion operators for integrable equations}
\author{\bf I.T. Habibullin and   A.R. Khakimova}

\begin{document}
\maketitle

\abstract{It is widely known that the recursion operator is a very important component of integrability. It allows one to describe in a compact form both hierarchies of the generalized symmetries and infinite series of the local conservation laws. In the literature, we can find several methods for constructing recursion operators, some of them use the Hamiltonian approach  and the others are based on the Lax representation of the equation. In the present article we discuss on an alternative method, suggested in \cite{HabKhaTMP18}, which is connected only with the first several generalized symmetries of the given equation. Efficiency of the method is illustrated with the examples of KdV, Krichever-Novikov and Kaup-Kupershmidt equations and discrete models.}

%\large
\normalsize

\section{Introduction}

In a series of our articles \cite{HabKhaPo}-\cite{HabKhaTMP18} we studied the properties and applications of the generalized invariant manifolds (GIM) for nonlinear equations. For a given differential (or discrete) equation GIM is a differential (or, respectively, difference) equation compatible with its linearization. We have shown that an appropriately chosen GIM generates the recursion operator, describing the hierarchy of the higher symmetries for the equation under consideration and also allows to construct the corresponding Lax pair. In the over mentioned articles for constructing the recursion operator we used a rather labor-consuming computational algorithm. In the present article we have significantly changed and improved the algorithm making it more convenient for use. 

The problem of constructing the recursion operators for the integrable equations has been investigated by many authors. Several methods are worked out to study the task. Some of them use the Lax representation (see, for instance, \cite{{ShabatIbragimov}}-\cite{Sokolov}). This way is very effective when the Lax pair is known. If it is not the case then it is reasonable to study directly the defining equation (see, equation (\ref{RtF}) below). To solve the equation the most authors use the multi-Hamiltonian approach (see \cite{{Gurses2}}-\cite{Oevel}). Their basic goal is to find two Hamiltonian operators $H_1$ and $H_2$ to the given equation. Then the recursion operator is given by the following formula $R=H_2H_1^{-1}$. However, usually the operators $H_1$ and $H_2$ are nonlocal and this causes some  difficulties in applying this method. 

In the present article we concentrate on the alternative method for constructing the recursion operator which is based on the symmetry approach (partly it is announced in \cite{HabKhaTMP18}). We assume that the recursion operator can be represented as a weakly nonlocal pseudo-differential operator of the form \cite{MaltsevNovikov} 
\begin{equation}\label{R_main}
R=R_0+\sum^{m}_{i=1}g^{(k_i)}D_x^{-1}h^{(i)}
\end{equation}
where $R_0$ is a differential operator. The non-local part consists of the combinations of the generators of the symmetries $u_{\tau_j}=g^{(j)}$ and the variational derivatives $h^{(j)}$ of the conserved densities (for more details see \cite{VladimirSokolov}). It was observed earlier
that in the most cases the recursion operator can be reduced to the form (\ref{R_main}). Note that the mentioned above symmetries $u_{\tau_j}=g^{(j)}$ are the members of the hierarchy of the symmetries started with the classical ones $g^{(1)}=u_x$, $g^{(2)}=u_t$ and followed by the higher symmetries arranged in ascending order
\begin{equation}\label{gm_sym}
g^{(1)}=u_x, \, g^{(2)}=u_t, \, g^{(3)}=u_{\tau_3}, \, \dots, \, g^{(m)}=u_{\tau_m},\dots ,
\end{equation}
such that the orders of the symmetries satisfy the inequalities $ord \,g^{(j)}< ord\, g^{(j+1)}$. We refer to the set $S=\{g^{(k_1)},g^{(k_2)},\dots, g^{(k_m)}\}$ of the generators in (\ref{R_main}) as the set of seed symmetries. In the simplest case $m=1$ exemplified by the KdV equation $S$ contains the only generator $g^{(1)}=u_x$. For the Kaup-Kupershmidt equation we have $m=2$ and $S$ contains two functions: $g^{(1)}=u_x$ and the right hand side $g^{(2)}=u_5+10uu_3+25u_1u_2+20u^2u_1$ of the KK equation (\ref{kk}) itself. For the Krichever-Novikov equation 
\begin{equation}\label{KN0}
u_t=u_{xxx}-\frac{3}{2}\frac{u_{xx}^2}{u_x}+\frac{P(u)}{u_x}\quad \mbox{with} \quad P'''''(u)=0
\end{equation}
there are two different recursion operators $R_1$ and $R_2$. For the first one we have $m=2$ and the seed symmetries are $g^{(1)}=u_x$ and $g^{(2)}=u_{xxx}-\frac{3}{2}\frac{u_{xx}^2}{u_x}+\frac{P(u)}{u_x}$. For the second recursion operator $m=3$ and $S$ contains in addition to these two generators also the generator of the fifth order symmetry of  the Krichever-Novikov equation 
$$g^{(3)}=u_5-5\frac{u_2u_4}{u_1}-\frac{5}{2}\frac{u_3^2}{u_1}+\frac{25}{2}\frac{u_3u_2^2}{u_1^2}-\frac{45}{8}\frac{u_2^4}{u_1^3}+\frac{5}{3}\left(\frac{5}{2}\frac{u_2^2}{u_1^3}-\frac{u_3}{u_1^2}\right)P(u)-\frac{5}{3}P'(u)\frac{u_2}{u_1}-\frac{5}{18}\frac{P^2(u)}{u_1^3}+\frac{5}{9}u_1P''(u)$$

Recall that for the Harry Dym equation $u_t=u^3u_{xxx}$ the recursion operator $R=u^3D_x^3uD_x^{-1}\frac{1}{u^2}$ found in \cite{Leo} is rewritten as follows 
$$R=u^4D_x^2\frac{1}{u^2}+3u^2u_1D_x\frac{1}{u^2}+3uu_2+u_tD_x^{-1}\frac{1}{u^2}.$$
Thus for the HD equation we have the only seed symmetry coinciding with the right hand side of the equation, $S=\left\{g^{(2)}=u^3u_3\right\}$, hence $m=1$.

Let us denote through $L_1$ a differential operator of the order $m$ which  annulates all of the functions $g^{(k_i)}$, $i=1,2...m$, in the set $S$. Theorem 1, presented below, asserts that then the operator
$ L_2 = L_1R $ is also a differential operator. Thus we come up to the following representation of the recursion operator as the ratio of two differential operators
\begin{equation}\label{ratio}
R=L_1^{-1}L_2
\end{equation} 
It is remarkable that this representation provides an effective tool for solving the defining equation (\ref{RtF}).

{\bf Remark}. Number $m$ of the non-localities in (\ref{R_main}) is closely connected with the action of the recursion operator on the hierarchy (\ref{gm_sym}), more precisely we have 
\begin{equation}\label{R-action}
Rg^{(j)}=g^{(j+m)}.
\end{equation} 
Therefore one can recover the whole hierarchy of the higher symmetries by the iterated application of $R$ to the set $S$.

Using the representation (\ref{ratio}), we can immediately find the Lax pair for the corresponding integrable equation (see below (\ref{psi-x}), (\ref{psi-t})), which, however, does not coincide with the usual.

The article is organized as follows. In \S2 we derive the formula (\ref{ratio}) and explain how to use it to solve effectively the defining equation (\ref{RtF}). The application of the symmetry algorithm for constructing the recursion operator is illustrated with the examples of the KdV, Kaup-Kupershmidt and Krichever-Novikov equations. In \S3 the algorithm is adopted to the integrable lattices. It is explained with the example of the Volterra chain. A rather complicated example of a non-autonomous lattice of the relativistic Toda type  with the periodic coefficients is studied in \S4. 

\section{Recursion operators for integrable PDE}

\subsection{Factorization of the recursion operators with weak non-localities}

Let us consider an integrable equation of the form
\begin{equation} \label{eq0}
u_t =f(u,u_1,u_2,...,u_k), \qquad \frac{\partial f}{\partial u_k}\neq0,
\end{equation}
where $u_j=D_x^{j} u$ and $D_x$ is the operator of the total derivative with respect to the variable $x$. The order $k$ of the highest order derivative of $u$ in the r.h.s. of the equation (\ref{eq0}) is called the order of the equation (\ref{eq0}). An evolutionary type equation of the order $r$
\begin{equation}\label{main_symmetry}
u_\tau=g(t,x,u,u_1, \dots, u_r)
\end{equation}
is called a symmetry of the equation (\ref{eq0}) if the flows defined by these two equations commute, i.e.
\begin{equation}\label{sym_eq0}
\left.D_tg-D_\tau f\right|_{(\ref{eq0}),(\ref{main_symmetry})}=0.
\end{equation}
It is widely known that any integrable equation (\ref{eq0}) possesses an infinite hierarchy of the symmetries which is completely described in a compact form in terms of the recursion operator $R$. 
In what follows we assume that equation (\ref{eq0}) admits the recursion operator of the form (\ref{R_main}).

{\bf Theorem 1.} For any weakly non-local pseudo-differential operator $R$ of the form (\ref{R_main}) there exists a pair of the differential operators $L_1$ and $L_2$ such that the following equation is satisfied
\begin{equation}\label{RconnectL1L2}
L_1R=L_2.
\end{equation}
{\bf Proof.} Let us define the differential operator
\begin{equation}\label{L1_general}
L_1=D_x^m+\alpha_1D_x^{m-1}+\dots+\alpha_m
\end{equation}
such that $L_1g^{(k_i)}=0$ for all $i=1,2,\dots,m.$ Evidently such kind operator exists. Now we have to show that $L_1R$ is also a differential operator. At first we show that the composition of the form $L_1g^{(k_i)}D_x^{-1}h^{(i)}$ is a differential operator. Indeed, due to the Leibniz rule we have
\begin{equation}\label{Dnonlocalcoeff}
D_x^jg^{(k_i)}D_x^{-1}h^{(i)}=D_x^j(g^{(k_i)})D_x^{-1}h^{(i)}+\sum^{j}_{k=1}c_j^kD_x^{j-k}(g^{(k_i)})D_x^{k-1}h^{(i)},
\end{equation}
where $c_j^k$, $k=1,2,\dots,j$ are the binominal coefficients. It is easily seen that only the first summand in (\ref{Dnonlocalcoeff}) contains the nonlocal term $D_x^{-1}h^{(i)}$. Therefore we can write 
\begin{equation}\label{Dnonlocalcoeff_2}
D_x^jg^{(k_i)}D_x^{-1}h^{(i)}=D_x^j(g^{(k_i)})D_x^{-1}h^{(i)}+H_{ij},
\end{equation}
where $H_{ij}$ for any $i$ and $j$ is a differential operator. Now we apply $L_1$ to $g^{(k_i)}D_x^{-1}h^{(i)}$ and due to the reasonings above we get
\begin{equation*}\label{L1nonlocalcoeff}
L_1g^{(k_i)}D_x^{-1}h^{(i)}=L_1(g^{(k_i)})D_x^{-1}h^{(i)}+\sum^{m}_{i,j=1}\alpha_{m-j}H_{ij}.
\end{equation*}
Since $L_1(g^{(k_i)})=0$ we find the relation 
\begin{equation*}\label{L1nonlocalcoeff_2}
L_1g^{(k_i)}D_x^{-1}h^{(i)}=\sum^{m}_{i,j=1}\alpha_{m-j}H_{ij}
\end{equation*}
which implies 
\begin{equation*}\label{L2_general}
L_2=L_1R_0+\sum^{m}_{i,j=1}\alpha_{m-j}H_{ij}.
\end{equation*}
Therefore $L_2$ is also a differential operator.

\subsection{Construction of the operators $L_1$, $L_2$ and $R$.}

In this section we assume that equation of the form (\ref{eq0}) admits the recursion operator (\ref{R_main}) and discuss the way to find it. 

Examples presented in the Introduction convince us that usually the set $S$ of the seed symmetries contains one or two classical symmetries $g^{(1)}=u_x$ and/or $g^{(2)}=u_t$ and only for the most complicated Krichever-Novikov equation the second recursion operator contains three generators $g^{(1)}=u_x$, $g^{(2)}=u_t$ and $g^{(3)}=u_{\tau}$. Hence when we look for the recursion operator we have to examine at least all of these possible choices of the set $S$: 
\begin{equation}\label{generalS}
S_1=\{u_x\},\quad S_2=\{u_t\},\quad S_3=\{u_x,\, u_t\},\quad S_4=\{u_x,\, u_t,\, u_{\tau}\},
\end{equation}
where $u_{\tau}=g^{(3)}$ is the closest higher symmetry of the equation (\ref{eq0}). 

For the chosen $S$ we take the operator $L_1=L_1(S)$ as the minimal order differential operator which annulates all of the generators in $S$. Obviously such an operator is not unique, it is defined up to multiplication by an arbitrary function. We can represent $L_1$ in an explicit form through the following determinant  
\begin{equation} \label{L1_det}
L_1U=\rho\left| \begin{array}{cccc}
D^{m}_xg^{(k_1)}&D^{m-1}_xg^{(k_1)}&\dots &g^{(k_1)} \\
\dots &\dots&\dots& \dots \\
D^{m}_xg^{(k_m)}&D^{m-1}_xg^{(k_m)}&\dots &g^{(k_m)} \\ 
D^{m}_xU&D^{m-1}_xU&\dots &U
\end{array} \right|,
\end{equation}
where $\rho$ is a function of the dynamical variables $u$, $u_1$, \dots . As soon as $L_1$ is found we can start to search $L_2$. As it follows from the Remark above (see Introduction) the order of the operator $L_2$ is determined as follows
\begin{equation}\label{diff_order}
m_2=m+ord\, g^{(k_1+m)}-ord\, g^{(k_1)}.
\end{equation}
In order to find $L_2$ we use the representation $R=L_1^{-1}L_2$. We substitute it into the defining equation for $R$ (see \cite{ShabatIbragimov}, \cite{Olver})
\begin{equation}\label{RtF}
\frac{d}{dt}R=\left[F^*,R\right].
\end{equation}
Here $F^*$ is the linearization operator of the equation (\ref{eq0}):
\begin{equation}\label{F*}
F^*=\left(\frac{\partial f}{\partial u}+\frac{\partial f}{\partial u_1}D_x+\frac{\partial f}{\partial u_2}D_x^2+...+\frac{\partial f}{\partial u_k}D_x^k\right).
\end{equation}
Let us substitute $R=L_1^{-1}L_2$ into the equation (\ref{RtF}) and after some simple transformations obtain 
\begin{equation}\label{AconnectL1L2}
\frac{d}{dt}(L_2)L_2^{-1}+L_2F^*L_2^{-1}=\frac{d}{dt}(L_1)L_1^{-1}+L_1F^*L_1^{-1}=:A.
\end{equation}
It immediately follows from (\ref{AconnectL1L2}) that the operators $L_1$ and $L_2$ solve one and the same equation
\begin{equation}\label{AconnectLj}
\frac{d}{dt}(L_j)=AL_j-L_jF^*, \quad j=1,2.
\end{equation}
It is easily checked that for any choice of $S$ the operator $A$ defined by (\ref{AconnectL1L2}) is a differential operator. Indeed, due to the construction the following equation \textsc{}
\begin{equation}\label{LU}
L_1U=0
\end{equation}
is compatible with the linearized equation
\begin{equation}\label{dtU}
\frac{d}{dt}U=F^*U.
\end{equation}
By applying the operator $\frac{d}{dt}$ to the equation (\ref{LU}) we obtain due to (\ref{dtU}) that
\begin{equation}\label{dtL}
\left(\frac{d}{dt}L_1+L_1F^*\right)U=0
\end{equation}
for any solution $U$ of the equation (\ref{LU}). Hence the kernels of the operators $L_1$ and $\frac{d}{dt}L_1+L_1F^*$ satisfy the relation $\ker L_1\subset \ker (\frac{d}{dt}L_1+L_1F^*)$. Therefore the latter operator is divided by the former one, i.e. there exists a differential operator $A$ such that $\frac{d}{dt}L_1+L_1F^*=AL_1$.

We begin with $S_1$, look for $A$ as discussed above. As soon as the operator $A$ is found we can use (\ref{AconnectLj}) with $j=2$ to construct operator $L_2$ of the order $m_2$.  If $L_2$ is found then we can define $R=L_1^{-1}L_2$, if such $L_2$ does not exist, we pass to the next choice of $S$. 

For the operators $L_1$ and $L_2$ found we can easily determine $R=L_1^{-1}L_2$. A convenient way to do this is to substitute the explicit expressions of the operators $L_1$, $L_2$ and the ansatz (\ref{R_main}) into the equation $L_1 R=L_2$ and then look for the unknown operator $R_0$ and the unknown coefficients $h^{(j)}$ in (\ref{R_main}) by comparison of the coefficients in front of the powers of $D_x$.

It is worth mentioning that the operators $L_1$ and $L_2$ allow one to derive the Lax pair for the equation (\ref{eq0}):
\begin{eqnarray}
&&L_2\psi=\lambda L_1 \psi, \label{psi-x} \\
&&\frac{d}{dt}\psi=F^*\psi \label{psi-t}
\end{eqnarray}
where $\lambda$ is the spectral parameter.
The Lax pair (\ref{psi-x}), (\ref{psi-t}) is not ``fake" since it is generated by the recursion operator.

\subsection{Examples.}

{\bf Example 1}. As an illustrative example we take the very well studied Korteweg-de Vries equation 
\begin{equation}\label{kdv}
u_t=u_3+uu_1.
\end{equation}
Among the potential values of the parameter $m$ we first choose $m=1$ and take $S=u_x$. Thus the first pretender for $L_1$ is as follows
\begin{equation} \label{kdv_L1_det}
L_1U=\rho\left| \begin{array}{cc}
u_{2} & u_1 \\ 
U_1 & U
\end{array} \right|.
\end{equation}
For the sake of simplicity we put $\rho=-1$, thus
\begin{equation}\label{kdv_L1}
L_1=u_1D_x-u_2.
\end{equation}
Direct computations show that the operator $A=\frac{d}{dt}(L_1)L_1^{-1}+L_1F^*L_1^{-1}$ where 
\begin{equation}\label{kdv_lin}
F^*=D_x^3+uD_x+u_1
\end{equation}
is a third order differential operator of the form
\begin{equation}\label{kdv_A}
A=D_x^3-\frac{3u_2}{u_1}D_x^2+\left(u+\frac{3u_2^2}{u_1^2}\right)D_x+3\left(\frac{u_4}{u_1}+u_1-\frac{u_2u_3}{u_1^2}\right).
\end{equation}
Due to the formula (\ref{diff_order}) the corresponding operator $L_2$ should be of the third order
\begin{equation}\label{kdv_L2_general}
L_2=\beta^{(3)}D_x^3+\beta^{(2)}D_x^2+\beta^{(1)}D_x+\beta^{(0)}.
\end{equation}
In order to find the unknown coefficients $\beta^{(i)}$ we substitute (\ref{kdv_lin}), (\ref{kdv_A}) and (\ref{kdv_L2_general}) into the equation
\begin{equation}\label{kdv_formAL2}
\frac{d}{dt}(L_2)=AL_2-L_2F^*,
\end{equation}
and by comparing the coefficients at the powers of $D_x$, we derive the equations for defining $\beta^{(3)}$, \dots, $\beta^{(0)}$. Omitting the computations we give only the answer:
\begin{equation}\label{kdv_L2}
L_2=u_1D_x^3-u_2D_x^2+\frac{2}{3}uu_1D_x+u_1^2-\frac{2}{3}uu_2.
\end{equation}
Since the requested $L_2$ is found we pass to the final stage. Having the operators $L_1$ and $L_2$ it is very easy to find the recursion operator $R=L_1^{-1}L_2$. Since $m=1$ we can conclude that $R$ has the only nonlocal term. Also we can evaluate the order of the operator $R_0$ -- differential part of $R$, it is the difference of $m_2$ and $m$ and equals two in this case. Therefore,
\begin{equation}\label{kdv_R_general}
R=r^{(2)}D_x^2+r^{(1)}D_x+r^{(0)}+u_1D_x^{-1}h.
\end{equation}
We substitute $L_1$, $L_2$ and $R$ into the equation $L_2=L_1R$ and get
\begin{eqnarray}\label{kdv_R_search}
&& u_1D_x^3-u_2D_x^2+\frac{2}{3}uu_1D_x+u_1^2-\frac{2}{3}uu_2=\nonumber\\
&& \qquad \left(u_1D_x-u_2\right)\left(r^{(2)}D_x^2+r^{(1)}D_x+r^{(0)}+u_1D_x^{-1}h\right).
\end{eqnarray}
From the latter we find $r^{(2)}=1$, $r^{(1)}=0$, $r^{(0)}=\frac{2}{3}u$, $h=1$ and therefore we have (see also \cite{Gardner})
\begin{equation}\label{kdv_R}
R=D_x^2+\frac{2}{3}u+\frac{1}{3}u_1D_x^{-1}.
\end{equation}
Recall that the usual representation of $R$ through the Hamiltonian operators $H_1=D_x$, $H_2=D_x^3+4uD_x+2u_1$ is as follows
\begin{equation}\label{kdv_hamilt}
R=H_2H_1^{-1}.
\end{equation}

{\bf Example 2}. As the second illustrative example we consider the Kaup-Kupershmidt equation
\begin{equation}\label{kk}
u_t=u_5+10uu_3+25u_1u_2+20u^2u_1.
\end{equation}
In what follows we will need in its linearization operator 
\begin{equation}\label{kk_lin}
F^*=D_x^5+10uD_x^3+25u_1D_x^2+\left(25u_2+20u^2\right)D_x+10u_3+40uu_1.
\end{equation}
In order to find the set $S$ of the seed symmetries for the equation (\ref{kk}) we have to examine the possible cases $S_1=\{u_x\}$, $S_2=\{u_t\}$, $S_3=\{u_x, u_t\}$, $S_4=\{u_x, u_t, u_{\tau}\}$, where $u_{\tau}=g^{(3)}$ is the next symmetry of the Kaup-Kupershmidt equation, it is of the seventh order.  

We started with the case $S=S_1$. For the corresponding $L_1$ we found $A$ from the equation (\ref{AconnectL1L2}), then since due to (\ref{diff_order}) $m_2=1+5-1=5$ we searched a fifth order differential operator $L_2$ satisfying the equation $\frac{d}{dt}L_2=AL_2-L_2F^*$ and observed that such operator does not exist. In a similar way we have verified that the case $S=S_2$ does not fit.

Then we passed to the case $S=S_3$ and succeeded. Operator $L_1$ is found from the relation
\begin{equation} \label{kk_L1_det}
L_1U=\left| \begin{array}{ccc}
u_3 & u_{2} & u_1 \\ 
u_{2t} & u_{1t} & u_{t}\\
U_2 & U_1 & U
\end{array} \right|
\end{equation}
and is of the form
\begin{equation}\label{kk_L1}
L_1=\alpha D_x^2+\beta D_x+\gamma,
\end{equation}
where 
\begin{eqnarray*}
&&\alpha=u_2u_5+10uu_2u_3-u_1u_6-35u_1^2u_3-10uu_1u_4-40uu_1^3,\\
&&\beta=10uu_1u_5-u_3u_5+45u_1^2u_4-10uu_3^2+120uu_1^2u_2+\\
&&\qquad 60u_1u_2u_3+40u_1^4+u_1u_7,\\
&&\gamma=-120uu_1u_2^2-10uu_2u_5+u_3u_6+35u_1u_3^2-60u_2^2u_3-\\
&&\qquad 40u_1^3u_2-u_2u_7+10uu_3u_4+40uu_1^2u_3-45u_1u_2u_4.
\end{eqnarray*}

Then we look for the operator 
\begin{equation}\label{kk_A_general}
A=\sum^{5}_{j=0}A^{(j)}D_x^j
\end{equation}
from the equation: 
\begin{equation}\label{kk_formAL1}
\frac{d}{dt}(L_1)=AL_1-L_1F^*.
\end{equation}
We note that the operators $A$ and $F^*$ always have one and the same order. We have solved the equation (\ref{kk_formAL1}) and found all of the coefficients $A^{(j)}$, however some of them turned out to be huge and hence we give in an explicit form only three of them:
\begin{eqnarray*}\label{kk_coeffA}
&&A^{(5)}=1, \qquad A^{(4)}=-5\frac{\alpha_x}{\alpha},\\
&&A^{(3)}=20\frac{\alpha_x^2}{\alpha^2}+5\frac{\beta\alpha_x}{\alpha^2}-5\frac{\beta_x}{\alpha}-10\frac{\alpha_{xx}}{\alpha}+10u.
\end{eqnarray*}
At the next step we look for the eighth order differential operator $L_2$ (recall that $m_2=m+ord\, g^{(k_1+m)}-ord\, g^{(k_1)}=2+ord\, g^{(3)}-ord\, g^{(1)}=2+7-1=8$
)
\begin{equation}\label{kk_L2_general}
L_2=\sum^{8}_{k=0}b^{(k)}D_x^k
\end{equation}
from the equation $\frac{d}{dt}(L_2)=AL_2-L_2F^*$. It turned out that such operator does exist. Omitting the computations we give only the answer:
\begin{eqnarray*}\label{kk_coeffL2}
&&b^{(8)}=\alpha, \quad b^{(7)}=\beta, \quad b^{(6)}=12u\alpha+\gamma, \quad b^{(5)}=60u_1\alpha+12u\beta, \\
&&b^{(4)}=(133u_2+36u^2)\alpha+48u_1\beta+12u\gamma, \\
&&b^{(3)}=(169u_3+264uu_1)\alpha+(85u_2+36u^2)\beta+36u_1\gamma, \\
&&b^{(2)}=(132u_4+394uu_2+381u_1^2+32u^3)\alpha+(84u_3+192uu_1)\beta+(49u_2+36u^2)\gamma, \\
&&b^{(1)}=(63u_5+304uu_3+852u_1u_2+240u^2u_1)\alpha\\
&&\qquad +(48u_4+202uu_2+189u_1^2+32u^3)\beta+(35u_3+120uu_1)\gamma, \\
&&b^{(0)}=(17u_6+122uu_4+444u_1u_3+324u_2^2+192u^2u_2+368uu_1^2)\alpha\\
&&\qquad +(15u_5+102uu_3+272u_1u_2+144u^2u_1)\beta+(13u_4+82uu_2+69u_1^2+32u^3)\gamma.
\end{eqnarray*}
We now find the required recursion operator $R$. We know that the order of its differential part is $6=m_2-m$ and it has two non-local terms, therefore it is of the form:
\begin{equation}\label{kk_R_general}
R=\sum^{6}_{j=0}r^{(j)}D_x^j+u_1D_x^{-1}h^{(1)}+u_tD_x^{-1}h^{(2)}.
\end{equation}
The relation $L_1R=L_2$ allows us to find all of the functional parameters in (\ref{kk_R_general}):
\begin{eqnarray*}\label{kk_coeffR}
&&r^{(6)}=1, \quad r^{(5)}=0, \quad r^{(4)}=12u, \quad r^{(3)}=36u_1, \quad r^{(2)}=49u_2+36u^2,\\
&&r^{(1)}=35u_3+120uu_1, \quad r^{(0)}=13u_4+82uu_2+69u_1^2+32u^3,\\
&&h^{(1)}=2u_2+8u^2, \quad h^{(2)}=2.
\end{eqnarray*}
So we have the final form of $R$ coinciding with that found earlier in \cite{{Gurses}}:
\begin{eqnarray}\label{kk_R}
R=D_x^6+12uD_x^4+36u_1D_x^3+\left(49u_2+36u^2\right)D_x^2+\left(35u_3+120uu_1\right)D_x\nonumber\\
\qquad +13u_4+82uu_2+69u_1^2+32u^3+u_1D_x^{-1}(2u_2+8u^2)+2u_tD_x^{-1}.
\end{eqnarray}
Having the operators $L_1$ and $L_2$ we can immediately write down the Lax pair of the form (\ref{psi-x}), (\ref{psi-t}), which does not coincide with the known one.

{\bf Example 3}. The next example is connected with the famous Krichever-Novikov equation \cite{Krichever-N} 
\begin{eqnarray}\label{kn}
u_t=u_{3}-\frac{3}{2}\frac{u_{2}^2}{u_1}+\frac{P(u)}{u_1}\quad \mbox{with} \quad P'''''(u)=0.
\end{eqnarray}
It is the most interesting representative of the class of the  third order integrable equations of the form (\ref{eq0}).  All the other equations of that class can be derived from (\ref{kn}) by appropriate manipulations like the limit procedure or the differential substitutions. The algebraic structures related to (\ref{kn}) are the most generic and complicated.
The linearization operator of the equation (\ref{kn}) is given by
\begin{eqnarray}\label{kn_lin}
F^*=D^3_{x}-\frac{3u_{2}}{u_1}D^2_{x}+\frac{3u_{2}^2-2P(u)}{2u_1^2}D_x+\frac{P'(u)}{u_1}.
\end{eqnarray}
The set $S$ for the equation (\ref{kn}) consists of the generators of two classical symmetries $g^{(1)}=u_x$ and $g^{(2)}=u_t$ hence $m=2$ and the operator $L_1$ is defined by
\begin{equation} \label{kn_L1_det}
L_1U=\left| \begin{array}{ccc}
u_3 & u_{2} & u_1 \\ 
u_{2t} & u_{1t} & u_{t}\\
U_2 & U_1 & U
\end{array} \right|.
\end{equation}
Coefficients of the operator $L_1$
\begin{equation}\label{kn_L1}
L_1=\alpha D_x^2+\beta D_x+\gamma,
\end{equation}
are as follows 
\begin{eqnarray*}
&&\alpha=4u_{2}u_{3}-u_1u_{4}-\frac{3u_2^3}{u_1}+\frac{2P(u)u_2}{u_1}-P'(u)u_1,\\
&&\beta=u_1u_5-3u_2u_4+\frac{9u_2^2u_3}{u_1}-4u_3^2-\frac{2P(u)(u_3u_1-u_2^2)}{u_1^2}-P'(u)u_2+P''(u)u_1^2-\frac{3u_2^4}{u_1^2},\\
&&\gamma=\left(u_3+\frac{u_2^2}{u_1}\right)P'(u)+\left(u_3+\frac{3u_2^2}{u_1}\right)u_4-u_2u_5-\frac{6u_2^3u_3}{u_1^2}-P''(u)u_1u_2-\frac{2Pu_2^3}{u_1^3}+\frac{3u_2^5}{u_1^3}.
\end{eqnarray*}
The corresponding operator $A$ is of the third order
\begin{equation}\label{kn_A_general}
A=\sum^{3}_{j=0}A^{(j)}D_x^j,
\end{equation}
it is found from the equation (\ref{AconnectLj}) with $j=1$
\begin{eqnarray*}\label{kn_coeffA}
&&A^{(3)}=1, \quad A^{(2)}=-3\frac{\alpha_x}{\alpha}-\frac{3u_{2}}{u_1},\\
&&A^{(1)}=-\frac{6u_{3}}{u_1}+\frac{6\alpha_x^2}{\alpha^2}+\frac{3(2\alpha u_2+\beta u_1)\alpha_x}{\alpha^2 u_1}+\frac{15 u_2^2}{2u_1^2}-\frac{3\alpha_{xx}}{\alpha}-\frac{3\beta_x}{\alpha}-\frac{P(u)}{u_1^2},\\
&&A^{(0)}=-\frac{3u_4}{u_1}+\frac{3(5\alpha u_2+\beta u_1+2\alpha_xu_1)u_3}{\alpha u_1^2}-\frac{\alpha_{xxx}}{\alpha}+\frac{3(\alpha u_2+2\alpha_xu_1+\beta u_1)\alpha_{xx}}{\alpha^2u_1}\\
&&\quad-\frac{6\alpha_x^3}{\alpha^3}-\frac{3(2\alpha u_2+3\beta u_1)\alpha_x^2}{\alpha^3u_1}+\left(\frac{9\beta_x}{\alpha^2}-\frac{3\beta^2}{\alpha^3}-\frac{6\beta u_2}{\alpha^2u_1}+\frac{P(u)}{\alpha u_1^2}-\frac{15u_2^2}{2\alpha u_1^2}+\frac{3\gamma}{\alpha^2}\right)\alpha_x\\
&&\quad -\frac{3\beta_{xx}}{\alpha}+\frac{3(2\alpha u_2+\beta u_1)\beta_x}{\alpha^2u_1}-\frac{P'(u)}{u_1}+\frac{\alpha_t}{\alpha}-\frac{3\gamma_x}{\alpha}-\frac{3\beta u_2^2}{\alpha u_1^2}+\frac{4P(u)u_2}{u_1^3}-\frac{12u_2^3}{u_1^3}.
\end{eqnarray*}
Let us evaluate the order of the operator $L_2$. According to the formula (\ref{diff_order}) we have $m_2=m+5-1=6$.  The coefficients of the operator $L_2$  
\begin{equation}\label{kn_L2_general}
L_2=\sum^{6}_{k=0}b^{(k)}D_x^k
\end{equation}
are found from (\ref{AconnectLj}) with $j=2$. 
They are of the form
\begin{eqnarray*}\label{kn_coeffL2}
&&b^{(6)}=\alpha, \quad b^{(5)}=\beta-\frac{4u_2}{u_1}\alpha, \quad
b^{(4)}=\left(\frac{14u_2^2}{u_1^2}-\frac{10u_3}{u_1}-\frac{4P(u)}{3u_1^2}\right)\alpha-\frac{4u_2}{u_1}\beta+\gamma, \\
&&b^{(3)}=\left(\frac{48u_2u_3}{u_1^2}-\frac{10u_4}{u_1}-\frac{38u_2^3}{u_1^3}+\frac{28P(u)u_2}{3u_1^3}-\frac{10P'(u)}{3u_1}\right)\alpha\\
&&\qquad \quad -\left(\frac{6u_3}{u_1}-\frac{10u_2^2}{u_1^2}+\frac{4P(u)}{3u_1^2}\right)\beta-\frac{4u_2}{u_1}\gamma, \\
&&b^{(2)}=\left(\frac{28u_3^2}{u_1^2}-\frac{5u_5}{u_1}+\frac{4P(u)^2}{9u_1^4}-\frac{14}{9}P''(u)+\frac{32P(u)u_3}{3u_1^3}+\frac{32P'(u)u_2}{3u_1^2}+\frac{32u_2u_4}{u_1^2}\right.\\
&&\qquad \quad\left.+\frac{69u_2^4}{u_1^4}-\frac{92P(u)u_2^2}{3u_1^4}-\frac{124u_2^2u_3}{u_1^3}\right)\alpha-\left(\frac{2u_3}{u_1}-\frac{6u_2^2}{u_1^2}+\frac{4P(u)}{3u_1^2}\right)\gamma\\
&&\qquad \quad-\left(\frac{4u_4}{u_1}-\frac{22u_2u_3}{u_1^2}+\frac{18u_2^3}{u_1^3}-\frac{20P(u)u_2}{3u_1^3}+\frac{2P'(u)}{u_1}\right)\beta,\\
&&b^{(1)}=\left(P'''(u)u_1-\frac{70u_2u_3^2}{u_1^3}+\frac{17P'(u)u_3}{3u_1^2}+\frac{8u_2u_5}{u_1^2}-\frac{8P(u)^2u_2}{3u_1^5}+\frac{48P(u)u_2^3}{u_1^5}\right.\\
&&\qquad \quad\left.+\frac{19u_3u_4}{u_1^2}+\frac{14P(u)u_4}{3u_1^3}-\frac{u_6}{u_1}-\frac{43u_2^2u_4}{u_1^3}+\frac{153u_2^3u_3}{u_1^4}-\frac{38P(u)u_2u_3}{u_1^4}-\frac{66u_2^5}{u_1^5}\right.\\
&&\qquad \quad\left.-\frac{14P'(u)u_2^2}{u_1^3}+\frac{4P(u)P'(u)}{3u_1^3}+\frac{P''(u)u_2}{u_1}\right)\alpha-\left(\frac{u_5}{u_1}-\frac{6u_2u_4}{u_1^2}+\frac{26u_2^2u_3}{u_1^3}\right.\\
&&\qquad \quad\left.-\frac{6u_3^2}{u_1^2}-\frac{15u_2^4}{u_1^4}+\frac{32P(u)u_2^2}{3u_1^4}-\frac{4P(u)u_3}{u_1^3}-\frac{2P'(u)u_2}{u_1^2}-\frac{4}{9}P''(u)-\frac{4P(u)^2}{9u_1^4}\right)\beta\\
&&\qquad \quad-\left(\frac{2u_4}{u_1}-\frac{8u_2u_3}{u_1^2}+\frac{6u_2^3}{u_1^3}-\frac{4P(u)u_2}{u_1^3}+\frac{2P'(u)}{3u_1}\right)\gamma,
\end{eqnarray*}
\begin{eqnarray*}
&&b^{(0)}=\left(\frac{4P(u)^2u_2^2}{3u_1^6}+\frac{23P'(u)u_2u_3}{3u_1^3}-\frac{2P(u)u_2u_4}{u_1^4}+\frac{4P(u)P'(u)u_2}{9u_1^4}-\frac{5P'(u)^2}{9u_1^2}\right.\\
&&\qquad \quad\left.+\frac{34P(u)u_2^2u_3}{3u_1^5}+\frac{4P(u)P''(u)}{9u_1^2}-\frac{7u_2u_3u_4}{u_1^3}-\frac{26P'(u)u_2^3}{3u_1^4}-\frac{14P(u)u_3^2}{3u_1^4}\right.\\
&&\qquad \quad\left.+\frac{10P''(u)u_2^2}{3u_1^2}-\frac{15u_2^4u_3}{u_1^5}-\frac{u_2^2u_5}{u_1^3}-\frac{4P(u)u_2^4}{u_1^6}-\frac{4P''(u)u_3}{3u_1}+\frac{2P(u)u_5}{3u_1^3}\right.\\
&&\qquad \quad\left.+\frac{3u_2^3u_4}{u_1^4}+\frac{2u_3u_5}{u_1^2}-\frac{5P'(u)u_4}{3u_1^2}+\frac{21u_2^2u_3^2}{u_1^4}-\frac{8P(u)^2u_3}{9u_1^5}-P'''(u)u_2\right.\\
&&\qquad \quad\left.-\frac{6u_3^3}{u_1^3}+\frac{3u_2^6}{u_1^6}+\frac{5}{9}P''''(u)u_1^2\right)\alpha+\left(\frac{5}{9}P'''(u)u_1-\frac{6u_2u_3^2}{u_1^3}-\frac{P'(u)u_3}{3u_1^2}\right.\\
&&\qquad \quad\left.+\frac{u_2u_5}{u_1^2}-\frac{8P(u)^2u_2}{9u_1^5}+\frac{u_3u_4}{u_1^2}+\frac{2P'(u)u_2^2}{3u_1^3}+\frac{4P(u)P'(u)}{9u_1^3}-\frac{P''(u)u_2}{u_1}\right.\\
&&\qquad \quad\left.+\frac{2P(u)u_4}{3u_1^3}-\frac{5u_2^2u_4}{u_1^3}+\frac{15u_2^3u_3}{u_1^4}-\frac{14P(u)u_2u_3}{3u_1^4}+\frac{16P(u)u_2^3}{3u_1^5}-\frac{6u_2^5}{u_1^5}\right)\beta\\
&&\qquad \quad+\left(\frac{u_5}{u_1}-\frac{4u_2u_4}{u_1^2}-\frac{2u_3^2}{u_1^2}+\frac{8u_2^2u_3}{u_1^3}-\frac{3u_2^4}{u_1^4}+\frac{4P(u)u_2^2}{3u_1^4}+\frac{4P(u)^2}{9u_1^4}\right.\\
&&\qquad \quad\left.-\frac{8P'(u)u_2}{3u_1^2}+\frac{10}{9}P''(u)\right)\gamma. \\
\end{eqnarray*}
Having the set $S$ we can easily write down the ansatz for the recursion operator $R$

\begin{equation}\label{kn_R_general}
R=\sum^{4}_{j=0}r^{(j)}D_x^j+u_1D_x^{-1}h^{(1)}+u_tD_x^{-1}h^{(2)}.
\end{equation}
The relation $L_1R=L_2$ allows one to find all of the functional parameters in (\ref{kn_R_general}):
\begin{eqnarray*}\label{kn_coeffR}
&&r^{(4)}=1, \quad r^{(3)}=-\frac{4u_2}{u_1}, \quad r^{(2)}=-\frac{2u_3}{u_1}+\frac{6u_2^2}{u_1^2}-\frac{4P(u)}{3u_1^2}, \\
&&r^{(1)}=-\frac{2u_4}{u_1}+\frac{8u_2u_3}{u_1^2}-\frac{6u_2^3}{u_1^3}+\frac{4P(u)u_2}{u_1^3}-\frac{2P'(u)}{3u_1}, \\
&&r^{(0)}=\frac{u_5}{u_1}-\frac{4u_2u_4}{u_1^2}-\frac{2u_3^2}{u_1^2}+\frac{8u_2^2u_3}{u_1^3}-\frac{3u_2^4}{u_1^4}+\frac{4P(u)u_2^2}{3u_1^4}\\
&&\qquad {} +\frac{4P(u)^2}{9u_1^4}-\frac{8P'(u)u_2}{3u_1^2}+\frac{10}{9}P''(u),\\
&&h^{(1)}=-\frac{u_6}{u_1^2}+\frac{6u_2u_5}{u_1^3}-\frac{5}{9}P'''(u)+\frac{5P''(u)u_2}{3u_1^2}-\frac{10P(u)^2u_2}{9u_1^6}\\
&&\qquad {} -\frac{10u_2(4u_1u_3-5u_2^2)P(u)}{3u_1^6}-\frac{15u_2(2u_1u_3-u_2^2)(2u_1u_3-3u_2^2)}{2u_1^6}\\
&&\qquad {} +\frac{5(2P(u)+12u_1u_3-27u_2^2)(3u_4+P'(u))}{18u_1^4}, \\
&&h^{(2)}=-\frac{u_4}{u_1^2}+\frac{4u_2u_3}{u_1^3}-\frac{3u_2^3}{u_1^4}-\frac{P'(u)}{u_1^2}+\frac{2P(u)u_2}{u_1^4}.
\end{eqnarray*}
Thus we obtain the final form of the operator $R$ 
\begin{eqnarray}\label{kn_R}
&&R=D_x^4-\frac{4u_2}{u_1}D_x^3+\left(\frac{6u_2^2}{u_1^2}-\frac{2u_3}{u_1}-\frac{4P(u)}{3u_1^2}\right)D_x^2\nonumber\\
&&\qquad +\left(\frac{8u_2u_3}{u_1^2}-\frac{2u_4}{u_1}-\frac{6u_2^3}{u_1^3}+\frac{4P(u)u_2}{u_1^3}-\frac{2P'(u)}{3u_1}\right)D_x+\frac{u_5}{u_1}-\frac{4u_2u_4}{u_1^2}\nonumber\\
&&\qquad -\frac{2u_3^2}{u_1^2}+\frac{8u_2^2u_3}{u_1^3}-\frac{3u_2^4}{u_1^4}+\frac{4P(u)u_2^2}{3u_1^4}+\frac{4P(u)^2}{9u_1^4}-\frac{8P'(u)u_2}{3u_1^2}+\frac{10}{9}P''(u)\nonumber\\
&&\qquad +u_1D_x^{-1}\left(\frac{6u_2u_5}{u_1^3}-\frac{u_6}{u_1^2}-\frac{5}{9}P'''(u)+\frac{5P''(u)u_2}{3u_1^2}-\frac{10P(u)^2u_2}{9u_1^6}\right.\nonumber\\
&&\qquad \left.-\frac{10u_2(4u_1u_3-5u_2^2)P(u)}{3u_1^6}-\frac{15u_2(2u_1u_3-u_2^2)(2u_1u_3-3u_2^2)}{2u_1^6}\right.\\
&&\qquad \left.+\frac{5(2P(u)+12u_1u_3-27u_2^2)(3u_4+P'(u))}{18u_1^4} \right)\nonumber\\
&&\qquad +u_tD_x^{-1}\left(\frac{4u_2u_3}{u_1^3}-\frac{u_4}{u_1^2}-\frac{3u_2^3}{u_1^4}-\frac{P'(u)}{u_1^2}+\frac{2P(u)u_2}{u_1^4}\right).\nonumber
\end{eqnarray}
It coincides with $R$ found earlier in \cite{Sokolov}.

\section{Construction of the recursion operators for the integrable lattices}

In this section we concentrate on the integrable lattices of the form
\begin{eqnarray}\label{discrete_main}
u_{n,t}=f(u_{n+k},u_{n+k-1},\dots,u_{n-k}), \quad \frac{\partial f}{\partial u_{n+k}}\frac{\partial f}{\partial u_{n-k}}\neq 0
\end{eqnarray}
where the sought function $u=u_n(t)$ depends on the integer $n$ and real $t$. The non-negative integer $k$ in (\ref{discrete_main}) is called the order of the equation (\ref{discrete_main}). A lattice of the same form
\begin{eqnarray}\label{discrete_main_symm}
u_{n,\tau}=g(u_{n+m},u_{n+m-1},\dots,u_{n-m})
\end{eqnarray}
is a symmery of the lattice (\ref{discrete_main}) if the following condition is satisfied
\begin{eqnarray}\label{cond_exist_symm}
\left.D_tg-D_\tau f\right|_{(\ref{discrete_main}),(\ref{discrete_main_symm})}=0.
\end{eqnarray}
Since the lattice (\ref{discrete_main}) is assumed to be integrable then it admits an infinite hierarchy of the symmetries 
\begin{equation}\label{dis-hier}
u_{\tau_1}=g^{(1)}, u_{\tau_2}=g^{(2)},\dots , u_{\tau_j}=g^{(j)},\dots .
\end{equation}
First several members of the hierarchy are the classical symmetries and the others are generalized ones. The hierarchy is effectively described by the recursion operator. In what follows we request that the recursion operator  $R$ is a pseudo-difference operator with weak non-localities, i.e. it can be reduced to the following form
\begin{eqnarray}\label{R_main_discrete}
R=R_0+\sum^{m}_{j=1}g^{(k_j)}(D_n-1)^{-1}h^{(j)}
\end{eqnarray}
where $D_n$ is the shift operator acting due to the rule $D_nq_n=q_{n+1}$, $R_0$ is a difference operator
\begin{eqnarray}\label{R0_main_discrete}
R_0=\gamma^{(s)}D_n^{s}+\gamma^{(s-1)}D_n^{s-1}+\dots+\gamma^{(-s)}D_n^{-s}, \quad s>0,
\end{eqnarray}
the coefficients $h^{(j)}$, $j=1,\dots,m$ are functions of the variables $u_n, u_{n\pm 1}, \dots $ and the set of coefficients $S=\left\{g^{(k_1)},g^{(k_2)},\dots ,g^{(k_m)}\right\}$ is a subset of the generators of the symmetries from the hierarchy (\ref{dis-hier}). It is supposed that the generators $\left\{g^{(k_1)},g^{(k_2)},\dots ,g^{(k_m)}\right\}$ constitute a linearly independent set over the field of the complex numbers.

{\bf Theorem 2.} Let $R$ be a weakly nonlocal difference operator of the form (\ref{R_main_discrete}). Then there exists a pair of the difference operators $L_1$ and $L_2$ of the form
\begin{eqnarray}\label{L1_dis_gen}
L_1=\alpha^{(0)}D_n^{m}+\alpha^{(1)}D_n^{m-1}+\dots+\alpha^{(m)},
\end{eqnarray}
\begin{eqnarray}\label{L2_dis_gen}
L_2=\beta^{(p)}D_n^{p}+\beta^{(p-1)}D_n^{p-1}+\dots+\beta^{(-q)}D_n^{-q},\quad p>m, \quad q>0
\end{eqnarray}
such that the following condition is satisfied
\begin{eqnarray}\label{L_1RL2_dis_gen}
L_1R=L_2.
\end{eqnarray}

{\bf Proof.} As $L_1$ we take the minimal order difference operator annulating all of the generators from the set $S$. In other words we should recover the difference operator (\ref{L1_dis_gen}) for the given fundamental system of its solutions $\left\{g^{(k_1)},g^{(k_2)},\dots ,g^{(k_m)}\right\}$.
Obviously the answer can be given in the following explicit form
\begin{equation} \label{L1_det_dis}
L_1U=\rho\left| \begin{array}{ccccc}
g^{(k_1)}_{m}&g^{(k_1)}_{m-1}&\dots &g^{(k_1)}_{1}& g^{(k_1)} \\
g^{(k_2)}_{m} &g^{(k_2)}_{m-1}&\dots &g^{(k_2)}_{1}& g^{(k_2)} \\
\dots &\dots&\dots& \dots & \dots\\
g^{(k_m)}_{m}&g^{(k_m)}_{m}&\dots &g^{(k_m)}_{1} &g^{(k_m)}\\ 
U_m&U_{m-1}&\dots &U_1 &U
\end{array} \right|,
\end{equation}
where $g_m^{(j)}:=D_n^m(g^{(j)})$, $U_j=D^j_nU$ and $\rho$ is an arbitrary function of the dynamical variables $u_n$, $u_{n\pm 1}$, \dots. 
Let us show that for this choice of $L_1$ the operator $L_2:=L_1R$ is also a difference operator, i.e. $L_2$ does not contain any nonlocal term. Actually, it is enough to prove that a product of the form $L_1g(D_n-1)^{-1}h$ is a difference operator whenever $L_1g=0$. Due to (\ref{L1_dis_gen}) this product can be written as follows 
\begin{eqnarray}\label{cond_T2}
&& g_m(D_n-1)^{-1}D_n^mh+\alpha^{(1)}g_{m-1}(D_n-1)^{-1}D_n^{m-1}h+\dots+\alpha^{(m)}(D_n-1)^{-1}h=\nonumber\\
&& \quad g_m(D_n-1)^{-1}(D_n^m-1)h+\alpha^{(1)}g_{m-1}(D_n-1)^{-1}(D_n^{m-1}-1)h\\
&& \qquad +\alpha^{(m-1)}h+L_1(g)(D_n-1)h,\nonumber
\end{eqnarray}
where $g_m:=D_n^m(g)$. Let us observe that the last term in (\ref{cond_T2}) vanishes since $L_1(g)=0$. All of the other terms in (\ref{cond_T2}) contain linear combinations of the products $(D_n-1)^{-1}(D_n^j-1)$ which do not produce nonlocalities for any integer $j\geq 0$.

For the discrete integrable equation (\ref{discrete_main}) the set $S$ is of one of the following forms
\begin{equation}\label{discrete-S}
S_1=\left\{g^{(0)}  \right\},\, S_2=\left\{f  \right\},\, S_3=\left\{g^{(0)},\, f  \right\},\, S_4=\left\{f,\,g^{(1)}  \right\},\, S_5=\left\{f,\,g^{(1)},\, g^{(2)}  \right\}. 
\end{equation}
Where $u_{\tau_0}=g^{(0)}$ is a zero order classical symmetry of the equation (\ref{discrete_main}), if the lattice admits such a symmetry, $u_t=f$ is the equation (\ref{discrete_main}) itself, $u_{\tau_1}=g^{(1)}$ and $u_{\tau_2}=g^{(2)}$ are the closest higher symmetries.  The set $S_1$ defines a set of seed symmetries for the coupled lattice discussed in \cite{HabKhaTMP18}, the case $S_2$  corresponds to the Volterra chain (see, example below), the set $S_3$ is related to the example studied below in the section 4, $S_4$ corresponds to the Yamilov discretization of the Krichever-Novikov equation (YdKN) and the case $S_5$ corresponds to the second recursion operator for the YdKN (see \cite{Mikh}, \cite{Wang}).

Let us assume now that the lattice (\ref{discrete_main}) admits a weakly non-local pseudo-difference recursion operator (\ref{R_main_discrete}) and discuss on the question how to find it. 

We have to determine the appropriate set of the seed symmetries. To this end we should examine the possible sets listed in (\ref{discrete-S}). Let us begin with the simplest one $S_1$ (or $S_2$ if the lattice does not admit any zero order symmetry). Then we determine the operator $L_1$ due to the determinant representation (\ref{L1_det_dis}) and look for the operator $L_2$ such that the ratio $R=L_1^{-1}L_2$ solves the defining equation for the recursion operator
\begin{equation}\label{R_eq_solv}
\frac{d}{dt}R=[F^*,R]
\end{equation}
where $F^*$ is the Frechet derivative (or the linearization operator) for the equation (\ref{discrete_main}):
\begin{equation}\label{frechet_der}
F^*=\frac{\partial f}{\partial u_{n+k}}D_n^{k}+\frac{\partial f}{\partial u_{n+k-1}}D_n^{k-1}+\dots+\frac{\partial f}{\partial u_{n-k}}D_n^{-k}.
\end{equation}
The natural numbers $m$, $p$ and $q$ from the formulas (\ref{L1_dis_gen}) and (\ref{L2_dis_gen}) are related to each other by the formulas
\begin{equation}\label{ord_L2_dis}
p=m+l, \quad q=l, \quad l=ord\, g^{(k_1+m)}-ord\, g^{(k_1)}.
\end{equation}

In order to construct the operator $L_2$ we use the scheme suggested above in section $2.2$. For the operator $L_1$ found in virtue of the formula (\ref{L1_det_dis}) we determine the operator $A$ from the equation
\begin{equation}\label{L1_form_dis}
\frac{d}{dt}L_1=AL_1-L_1F^{*}.
\end{equation}
In the next stage we examine the equation
\begin{equation}\label{L2_form_dis}
\frac{d}{dt}L_2=AL_2-L_2F^{*}
\end{equation}
and clarify whether it has a solution $L_2$ of the form (\ref{L2_dis_gen}) with the parameters $p$ and $q$ given by (\ref{ord_L2_dis}). If the answer is positive then we can conclude that $R=L_1^{-1}L_2$ is the recursion operator for the lattice  (\ref{discrete_main}). Otherwise we pass to the next member of the sequence (\ref{discrete-S}) and so on.

As it follows from the reasonings above (see Theorem 2) if the lattice (\ref{discrete_main}) admits a weakly nonlocal recursion operator then it can be found by using this procedure. In other words for some choice of the set of seed symmetries the operator $L_2$ exists satisfying the equation (\ref{L2_form_dis}). Assume that such $L_2$ is found, then to specify the undetermined coefficients in (\ref{R_main_discrete}) we use the equation $L_2=L_1R$  written in an enlarged form 
\begin{eqnarray}\label{search_coeff_R}
&& \beta^{(p)}D_n^{p}+\beta^{(p-1)}D_n^{p-1}+\dots+\beta^{(-q)}D_n^{-q}=\nonumber\\
&& \,(\alpha^{(0)}D_n^{m}+\alpha^{(1)}D_n^{m-1}+\dots+\alpha^{(m)})(\gamma^{(s)}D_n^{s}+\dots+\gamma^{(-s)}D_n^{-s})+\nonumber\\
&& \quad (\alpha^{(0)}D_n^{m}+\alpha^{(1)}D_n^{m-1}+\dots+\alpha^{(m)})\times\\
&& \qquad (g^{(1)}(D_n-1)^{-1}h^{(1)}+\dots+g^{(m)}(D_n-1)^{-1}h^{(m)}).\nonumber
\end{eqnarray}
Actually the comparison of the coefficients in front of the different powers of $D_n$ in (\ref{search_coeff_R}) allows one to find effectively the coefficients $h^{(j)}$ and $\gamma^{(j)}$.

\subsection{Evaluation of the recursion operator for the Volterra chain}

As an illustrative example of the application of the algorithm above we take the well-known Volterra chain
\begin{equation}\label{Volterra}
\frac{d}{dt}u_n=u_n(u_{n+1}-u_{n-1}).
\end{equation}
Let us find its linearization
\begin{equation}\label{Volterra_lin}
\frac{d}{dt}U_n=F^*U_n, \quad \mbox{where} \quad F^*=u_nD_n+(u_{n+1}-u_{n-1})-u_nD_n^{-1}.
\end{equation}
Since the Volterra chain doesn't admit any zero order autonomous symmetry we start with the set $S_2$. Let us define the operator $L_1$ through the determinant representation 
(\ref{L1_det_dis})
\begin{equation} \label{Volterra_L1_det}
L_1U_n=-\frac{1}{u_{n,t}u_{n+1,t}}\left| \begin{array}{cc} 
u_{n+1,t} & u_{n,t}\\
U_{n+1} & U_n
\end{array} \right|
\end{equation}
or, the same thing
\begin{equation}\label{Volterra_L1}
L_1=(D_n-1)\frac{1}{u_{n,t}}.
\end{equation}
Then we look for the operator $A$ of the form 
\begin{equation}\label{Volterra_A}
A=A^{(1)}D_n+A^{(0)}+A^{(-1)}D_n^{-1}
\end{equation}
from the equation (\ref{L1_form_dis}):
\begin{eqnarray}\label{Volterra_searchA}
&&\frac{1}{(u_{n+1,t})_t}D_n-\frac{1}{(u_{n,t})_t}=\left(A^{(1)}D_n+A^{(0)}+A^{(-1)}D_n^{-1}\right)\left(\frac{1}{u_{n+1,t}}D_n-\frac{1}{u_{n,t}}\right)\nonumber\\
&&\quad-\left(\frac{1}{u_{n+1,t}}D_n-\frac{1}{u_{n,t}}\right)\left(u_nD_n+(u_{n+1}-u_{n-1})-u_nD_n^{-1}\right).
\end{eqnarray}
By comparison of the coefficients in front of the powers of the operator $D_n$ in the last equation we obtain
\begin{equation}\label{Volterra_coeffA}
A^{(1)}=\frac{u_{n+1}u_{n+2,t}}{u_{n+1,t}}, \,
A^{(0)}=\frac{u_{n+1}u_{n,t}}{u_{n+1,t}}-\frac{u_nu_{n+1,t}}{u_{n,t}}, \,
A^{(-1)}=-\frac{u_nu_{n-1,t}}{u_{n,t}}.
\end{equation}
Due to the conjecture (\ref{ord_L2_dis}) the corresponding operator $L_2$ should be of the form 
\begin{equation}\label{Volterra_L2}
L_2=b^{(2)}D_n^2+b^{(1)}D_n+b^{(0)}+b^{(-1)}D_n^{-1}.
\end{equation}
Let us substitute the ansatz (\ref{Volterra_L2}) into the equation (\ref{L2_form_dis}) and find
\begin{eqnarray}\label{Volterra_searchL2}
&&b_t^{(2)}D_n^2+b_t^{(1)}D_n+b_t^{(0)}+b_t^{(-1)}D_n^{-1}=\nonumber\\
&&\left(A^{(1)}D_n+A^{(0)}+A^{(-1)}D_n^{-1}\right)\left(b^{(2)}D_n^2+b^{(1)}D_n+b^{(0)}+b^{(-1)}D_n^{-1}\right)\\
&&-\left(b^{(2)}D_n^2+b^{(1)}D_n+b^{(0)}+b^{(-1)}D_n^{-1}\right)\left(u_nD_n+(u_{n+1}-u_{n-1})-u_nD_n^{-1}\right).\nonumber
\end{eqnarray}
By collecting the coefficients at the diferent powers of $D_n$ we can derive some equations for the factors $b^{(j)}$. We omit the details and give only the answers
\begin{eqnarray}\label{Volterra_coeffL2}
b^{(2)}=\frac{c_1}{u_{n+2}-u_n}, \quad
b^{(1)}=\frac{c_3+c_1(u_{n+2}+u_{n+1})}{u_{n+1}(u_{n+2}-u_n)}-\frac{c_1}{u_{n+1}-u_{n-1}}, \nonumber\\
b^{(0)}=\frac{c_1}{u_{n+2}-u_n}-\frac{c_3+c_1(u_{n}+u_{n-1})}{u_n(u_{n+1}-u_{n-1})}, \quad
b^{(-1)}=-\frac{c_1}{u_{n+1}-u_{n-1}}.
\end{eqnarray}
Without loss of generality we set $c_1=1$ and $c_3=0$ and obtain
\begin{eqnarray}\label{L2}
L_2=\frac{1}{u_{n+2}-u_n}D_n^2+\left(\frac{u_{n+2}+u_{n+1}}{u_{n+1}(u_{n+2}-u_n)}-\frac{1}{u_{n+1}-u_{n-1}}\right)D_n +\\
\frac{1}{u_{n+2}-u_n}-\frac{u_{n}+u_{n-1}}{u_n(u_{n+1}-u_{n-1})}-\frac{1}{u_{n+1}-u_{n-1}}D_n^{-1}.\nonumber
\end{eqnarray}
Thus we have both operators $L_1$ and $L_2$ and are ready to construct the recursion operator $R$:
\begin{equation}\label{Volterra_R}
R=r^{(1)}D_n+r^{(0)}+r^{(-1)}D_n^{-1}+u_{n,t}(D_n-1)^{-1}h.
\end{equation}
Let us substitute into the equation $L_2=L_1R$ the detailed representations of the factors given by (\ref{Volterra_L1}), (\ref{Volterra_L2}), (\ref{Volterra_R}). As a result we get
\begin{eqnarray}\label{Volterra_searchR}
&&b^{(2)}D_n^2+b^{(1)}D_n+b^{(0)}+b^{(-1)}D_n^{-1}=\\
&&\left(\frac{1}{u_{n+1,t}}D_n-\frac{1}{u_{n,t}}\right)\left(r^{(1)}D_n+r^{(0)}+r^{(-1)}D_n^{-1}+u_{n,t}(D_n-1)^{-1}h\right),\nonumber
\end{eqnarray}
which is easily solved
\begin{equation}\label{Volterra_coeffR}
r^{(1)}=u_n, \quad
r^{(0)}=(u_{n+1}+u_n), \quad
r^{(-1)}=u_n, \quad
h=\frac{1}{u_n}.
\end{equation}
Therefore the recursion operator reads as follows 
\begin{equation}\label{Volterra_Rfinal}
R=u_nD_n+(u_{n+1}+u_n)+u_nD_n^{-1}+u_{n,t}(D_n-1)^{-1}\frac{1}{u_n}.
\end{equation}
It coincides with $R$ found earlier in \cite{ShabatIbragimov}, \cite{Zhang}. Note that in \cite{Zhang} this operator is represented as a ratio $R=H_2H_1^{-1}$ of two Hamiltonian operators 
$$H_1=u_n(D-D^{-1})u_n, \quad H_2=u_n(Du_nD+u_nD+Du_n-u_nD^{-1}- D^{-1}u_n-D^{-1}u_nD^{-1})u_n.$$

\section{A non-autonomous example}
Nowadays non-autonomous integrable lattices are intensively studied. Symmetry integrability conditions for this class of equations has been formulated in \cite{Levi}. The symmetry algorithm for constructing the recursion operator can be effectively applied to the non-autonomous lattices as well. Let us consider the following non-autonomous lattice of the relativistic Toda type
\begin{equation}\label{nonaut_eq}
u_{n,t}=h_nh_{n-1}(a_nu_{n+2}-a_{n-1}u_{n-2})
\end{equation}
where $h_n=u_{n+1}u_n-1$ and the coefficient $a_n$ is an arbitrary periodic function of the period 2, $a_{n+2}=a_n$. 
This lattice has been found within the frame of the integrable classification of the discrete equations on the quadratic graphs. Namely, it is a symmetry in the direction $n$ of the equation
\begin{equation}\label{quadeq}
u_{n+1,m+1}(u_{n,m}-u_{n,m+1})-u_{n+1,m}(u_{n,m}+u_{n,m+1})+2=0.
\end{equation}
Notice that equations (\ref{nonaut_eq}) and (\ref{quadeq}) were found by Garifullin and Yamilov in \cite{GY12}. In \cite{GYM14} the recursion operator for the non-autonomous lattice (\ref{nonaut_eq}) is constructed in an unusual way. The authors observed that the lattice (\ref{nonaut_eq}) can be reduced to an autonomous system found by Tsuchida in \cite{Tsuchida}, for which the recursion operator has already been found in \cite{Mikh}. In \cite{GYM14} the known recursion operator was recalculated to the scalar form in an appropriate way.

In this article, we derive the recursion operator directly using the symmetry algorithm discussed above.

As it has been observed in \cite{GYM14} the lattice (\ref{nonaut_eq}) possesses a rather large hierarchy of the symmetries. Since the lattice is  not autonomous then the set S of the seed symmetries obviously might also contain non-autonomous symmetries. The first two members of the symmetry hierarchy are as follows 
\begin{eqnarray}
u_{n,\tau_1}=(-1)^nu_n\label{nonaut_eq_sym_1},\\
u_{n,\tau_2}=h_nh_{n-1}(c_nu_{n+2}-c_{n-1}u_{n-2}), \quad c_{n+2}=c_n,\label{nonaut_eq_sym_2}
\end{eqnarray}
where $c_n$ is an arbitrary periodic function of $n$ with period equal to two.

As potential sets of seed symmetries, consider the following three sets:
\begin{equation}\label{nonaut_set}
S_1=\left\{u_{n,\tau_1}\right\},\quad S_2=\left\{u_{n,\tau_2}\right\},\quad S_3=\left\{u_{n,\tau_1};u_{n,\tau_2} \right\}.
\end{equation}
We checked that the first two sets do not fit, but the latest is surely the required set of seed symmetries.

The operator $L_1$ corresponding to $S_3$ is given by
\begin{equation} \label{nonaut_eq_L1_det}
L_1U_n=\left| \begin{array}{ccc}
D_n^2(u_{n,\tau_1}) & D_n(u_{n,\tau_1}) & u_{n,\tau_1}\\ 
D_n^2(u_{n,\tau_2}) & D_n(u_{n,\tau_2}) & u_{n,\tau_2}\\ 
U_{n+2} & U_{n+1} & U_n
\end{array} \right|
\end{equation}
As a result we have
\begin{equation}\label{nonaut_eq_L1}
L_1=\alpha D_n^2+\beta D_n+\gamma,
\end{equation}
where
\begin{eqnarray*}
&&\alpha=(-1)^{n+1}h_n\left(u_{n+1}h_{n-1}(c_nu_{n+2}-c_{n-1}u_{n-2})+u_{n}h_{n+1}(c_{n-1}u_{n+3}-c_{n}u_{n-1})\right),\\
&&\beta=(-1)^{n+1}\left(u_{n+2}h_nh_{n-1}(c_nu_{n+2}-c_{n-1}u_{n-2})-u_{n}h_{n+1}h_{n+2}(c_{n}u_{n+4}-c_{n-1}u_{n})\right),\\
&&\gamma=(-1)^{n}h_{n+1}\left(u_{n+1}h_{n+2}(c_nu_{n+4}-c_{n-1}u_{n})+u_{n+2}h_{n}(c_{n-1}u_{n+3}-c_{n}u_{n-1})\right).
\end{eqnarray*}
Obviously the linearization of (\ref{nonaut_eq}) is
\begin{equation}\label{nonaut_eq_lin}
U_{n,t}=F^*U_n,
\end{equation}
with 
\begin{eqnarray*}
&&F^*=a_nh_nh_{n-1}D_n^2+u_nh_{n-1}(a_nu_{n+2}-a_{n-1}u_{n-2})D_n+\\
&&\quad (u_{n+1}h_{n-1}+u_{n-1}h_{n})(a_nu_{n+2}-a_{n-1}u_{n-2})+\\
&&\qquad u_nh_{n}(a_nu_{n+2}-a_{n-1}u_{n-2})D_n^{-1}-a_{n-1}h_nh_{n-1}D_n^{-2}.
\end{eqnarray*}
In order to find the operator
\begin{equation}\label{nonaut_eq_A}
A=A^{(2)}D_n^2+A^{(1)}D_n+A^{(0)}+A^{(-1)}D_n^{-1}+A^{(-2)}D_n^{-2}.
\end{equation}
We solve the equation
\begin{equation}\label{L1_form_dis2}
\frac{d}{dt}L_1=AL_1-L_1F^{*},
\end{equation}
which is equivalent to 
\begin{eqnarray}\label{nonaut_eq_searchA}
&&\alpha_t D_n^2+\beta_t D_n+\gamma_t=\nonumber\\
&&\quad(A^{(2)}D_n^2+A^{(1)}D_n+A^{(0)}+A^{(-1)}D_n^{-1}+A^{(-2)}D_n^{-2})(\alpha D_n^2+\beta D_n+\gamma)\nonumber\\
&&\qquad-(\alpha D_n^2+\beta D_n+\gamma)\left(a_nh_nh_{n-1}D_n^2+u_nh_{n-1}(a_nu_{n+2}-a_{n-1}u_{n-2})D_n\right.\nonumber\\
&&\qquad {} \quad {}\left.+(u_{n+1}h_{n-1}+u_{n-1}h_{n})(a_nu_{n+2}-a_{n-1}u_{n-2})\right.\\
&&\qquad {} \qquad {}\left.+u_nh_{n}(a_nu_{n+2}-a_{n-1}u_{n-2})D_n^{-1}-a_{n-1}h_nh_{n-1}D_n^{-2}\right).\nonumber
\end{eqnarray}
It is easily checked that (\ref{nonaut_eq_searchA}) implies
\begin{eqnarray}\label{nonaut_eq_coeffA}
&&A^{(2)}=\frac{a_nh_{n+1}h_{n+2}(u_{n+1}g_n+u_{n}g_{n+1})}{u_{n+3}g_{n+2}+u_{n+2}g_{n+3}},\nonumber\\
&&A^{(1)}=\frac{a_{n-1}h_{n+1}(u_nh_ng_{n+2}+u_n^2u_{n+2}g_{n+1}+u_{n+2}g_{n})}{u_{n+2}g_{n+1}+u_{n+1}g_{n+2}}\nonumber\\
&& \qquad -\frac{a_{n}h_{n+1}(u_{n+1}g_{n}+u_{n}g_{n+1})(u_{n+4}g_{n+2}+u_{n+2}^2u_{n+4}g_{n+3}+u_{n+2}h_{n+2}g_{n+4})}{(u_{n+2}g_{n+1}+u_{n+1}g_{n+2})(u_{n+3}g_{n+2}+u_{n+2}g_{n+3})},\nonumber\\
&&A^{(0)}=\frac{d}{d t}\ln(u_{n+1}g_n+u_{n}g_{n+1})+ \left(\frac{u_{n+1}u_{n+3}h_{n}(u_{n+2}g_{n}-u_{n}g_{n+2})}{u_{n+1}g_n+u_{n}g_{n+1}}\right.\\
&& \qquad \left.-\frac{h_{n+1}(u_{n+2}g_{n}-u_{n}g_{n+2})(u_{n+1}g_{n+3}-u_{n+3}g_{n+1})}{(u_{n+1}g_n+u_{n}g_{n+1})(u_{n+2}g_{n+1}+u_{n+1}g_{n+2})}\right.\nonumber\\
&& \qquad \left.-\frac{u_n^2h_{n+1}(u_{n+1}g_{n+3}-u_{n+3}g_{n+1})}{u_{n+1}g_n+u_{n}g_{n+1}}-u_n(u_{n+3}h_{n+1}+u_{n+1}h_{n+2})\right)a_{n-1}\nonumber\\
&&  \qquad +\left(\frac{h_{n+1}(u_{n+1}g_{n+3}-u_{n+3}g_{n+1})(u_{n+4}g_{n+2}+u_{n+4}u_{n+2}^2g_{n+3}+u_{n+2}h_{n+2}g_{n+4})}{(u_{n+3}g_{n+2}+u_{n+2}g_{n+3})(u_{n+2}g_{n+1}+u_{n+1}g_{n+2})}\right. \nonumber\\
&&  \qquad \left.-\frac{u_{n+1}u_{n-1}h_{n}(u_{n+2}g_{n}-u_{n}g_{n+2})}{u_{n+1}g_n+u_{n}g_{n+1}}-\frac{h_{n-1}h_{n}(u_{n+2}g_{n+1}+u_{n+1}g_{n+2})}{u_{n+1}g_n+u_{n}g_{n+1}}\right.\nonumber\\
&& \qquad \left.+\frac{h_{n+1}h_{n+2}(u_{n+3}g_{n+4}+u_{n+4}g_{n+3})}{u_{n+3}g_{n+2}+u_{n+2}g_{n+3}}+u_{n+4}(u_{n+3}h_{n+1}+u_{n+1}h_{n+2})\right)a_n\nonumber\\
&&A^{(-1)}=-\frac{a_{n}h_{n}(u_ng_{n+2}+u_{n+2}h_{n+1}g_{n}+u_nu_{n+2}^2g_{n+1})}{u_{n+1}g_{n}+u_{n}g_{n+1}}\nonumber\\
&& \qquad +\frac{a_{n-1}h_{n}(u_{n+2}g_{n+1}+u_{n+1}g_{n+2})(u_{n-2}g_{n}+u_{n}h_{n-1}g_{n-2}+u_{n-2}u_{n}^2g_{n-1})}{(u_{n+1}g_{n}+u_{n}g_{n+1})(u_{n-1}g_{n}+u_{n}g_{n-1})},\nonumber\\
&&A^{(-2)}=-\frac{a_{n-1}h_{n}h_{n-1}(u_{n+2}g_{n+1}+u_{n+1}g_{n+2})}{u_{n-1}g_{n}+u_{n}g_{n-1}},\nonumber
\end{eqnarray}
where $g_n=h_nh_{n-1}(c_nu_{n+2}-c_{n-1}u_{n-2})$.

Next we look for the operator $L_2$ which has to be of the form
\begin{equation}\label{nonaut_eq_L2}
L_2=b^{(4)}D_n^4+b^{(3)}D_n^3+b^{(2)}D_n^2+b^{(1)}D_n+b^{(0)}+b^{(-1)}D_n^{-1}+b^{(-2)}D_n^{-2}.
\end{equation}
Indeed in (\ref{L2_dis_gen}) due to (\ref{ord_L2_dis}) we have $p=m+l$, $q=l$ where $m=2$, $l=2$.

Therefore the defining equation
\begin{equation*}
\frac{d}{dt}L_2=AL_2-L_2F^{*}
\end{equation*}
takes the form:
\begin{eqnarray}\label{nonaut_eq_searchL2}
&&b^{(4)}_tD_n^4+b^{(3)}_tD_n^3+b^{(2)}_tD_n^2+b^{(1)}_tD_n+b^{(0)}_t+b^{(-1)}_tD_n^{-1}+b^{(-2)}_tD_n^{-2}=\nonumber\\
&& \, \left(A^{(2)}D_n^2+A^{(1)}D_n+A^{(0)}+A^{(-1)}D_n^{-1}+A^{(-2)}D_n^{-2}\right)\times\nonumber\\
&&\quad \left(b^{(4)}D_n^4+b^{(3)}D_n^3+b^{(2)}D_n^2+b^{(1)}D_n+b^{(0)}+b^{(-1)}D_n^{-1}+b^{(-2)}D_n^{-2}\right)\\
&&\quad -\left(b^{(4)}D_n^4+b^{(3)}D_n^3+b^{(2)}D_n^2+b^{(1)}D_n+b^{(0)}+b^{(-1)}D_n^{-1}+b^{(-2)}D_n^{-2}\right)\times\nonumber\\
&&\qquad \left(a_nh_nh_{n-1}D_n^2+u_nh_{n-1}(a_nu_{n+2}-a_{n-1}u_{n-2})D_n\right.\nonumber\\
&&\qquad {} \quad {} \left.+(u_{n+1}h_{n-1}+u_{n-1}h_{n})(a_nu_{n+2}-a_{n-1}u_{n-2})\right.\nonumber\\
&&\qquad {} \qquad {} \left.+u_nh_{n}(a_nu_{n+2}-a_{n-1}u_{n-2})D_n^{-1}-a_{n-1}h_nh_{n-1}D_n^{-2}\right).\nonumber
\end{eqnarray}
By comparison of the coefficients in front of the powers of $D_n$ we get several equations which imply 
\begin{eqnarray*}\label{nonaut_eq_coeffL2}
&&b^{(4)}=(-1)^{n+1}\mu c_nh_{n+1}h_{n+2}(u_{n+1}g_n+u_ng_{n+1}),\nonumber\\
&&b^{(3)}=(-1)^{n+1}\mu c_{n-1}h_{n}h_{n+1}(u_{n+2}g_n-u_ng_{n+2})+(-1)^{n+1}\mu\frac{u_{n+2}g_{n+2}(u_{n+1}g_n+u_ng_{n+1})}{h_{n+2}},\nonumber\\
&&b^{(2)}=(-1)^{n+1}\mu (u_{n+1}g_n+u_ng_{n+1}) \left(\frac{u_{n+1}g_{n+2}-u_{n+2}g_{n+1}}{h_{n+1}}+c_n(h_{n+1}h_{n+3}-1)+\right.\nonumber\\
&& \qquad {} \left.c_{n-1}(h_{n}h_{n+2}-1)\right)+(-1)^{n+1}\mu\frac{u_{n+1}g_{n+1}(u_{n+2}g_n-u_ng_{n+2})}{h_{n+1}}+\nonumber\\
&& \qquad {} (-1)^n\mu c_{n}h_{n}h_{n-1}(u_{n+1}g_{n+2}+u_{n+2}g_{n+1})+(s_n^{(1)}+(-1)^ns_n^{(2)})(u_{n+1}g_n+u_ng_{n+1})\nonumber\\
&&b^{(1)}=(-1)^{n+1}\mu (u_{n+2}g_n-u_ng_{n+2}) \left(\frac{u_{n}g_{n+1}-u_{n+1}g_{n}}{h_{n}}+c_n(h_{n-1}h_{n+1}-1)+\right.\nonumber\\
&& \qquad {} \left.c_{n-1}(h_{n}h_{n+2}-1) \right)+(-1)^{n}\mu\frac{u_{n}g_{n}(u_{n+1}g_{n+2}+u_{n+2}g_{n+1})}{h_{n}}+\nonumber\\
&& \qquad {} (-1)^{n+1}\mu (u_{n+1}g_n+u_ng_{n+1})\frac{u_nh_{n+1}g_{n+2}+u_{n+2}h_ng_{n+2}+u_{n+2}h_{n+1}g_n}{h_{n}h_{n+1}}-\nonumber\\
&& \qquad {} (s_n^{(1)}+(-1)^ns_n^{(2)})(u_{n}g_{n+2}-u_{n+2}g_{n}),\nonumber\\
&&b^{(0)}=(-1)^n\mu (u_{n}g_{n+2}-u_{n+2}g_{n})\left(\frac{u_{n+1}g_{n+1}}{h_n}+\frac{u_{n-1}g_{n+1}+u_{n+1}g_{n-1}}{h_{n-1}}\right)+\nonumber\\
&& \qquad {} (-1)^{n+1}\mu (u_{n+1}g_n+u_ng_{n+1})\left(c_{n-1}h_{n+1}h_{n+2}+\frac{u_{n+1}g_{n+2}-u_{n+2}g_{n+1}}{h_n}+\right.\nonumber\\
&& \qquad {} \left.\frac{u_{n-1}g_{n+2}-u_{n+2}g_{n-1}}{h_{n-1}}\right)+(-1)^n\mu (u_{n+1}g_{n+2}+u_{n+2}g_{n+1})\left(\frac{u_{n-1}g_{n}-u_{n}g_{n-1}}{h_{n-1}}+\right.\nonumber\\
&& \qquad {} \left.c_n(h_{n-1}h_{n+1}-1)+c_{n-1}(h_{n}h_{n-2}-1)\right)-(s_n^{(1)}+(-1)^ns_n^{(2)})(u_{n+1}g_{n+2}+u_{n+2}g_{n+1})\nonumber\\
&&b^{(-1)}=(-1)^{n+1}\mu c_{n}h_{n}h_{n+1}(u_{n+2}g_n-u_ng_{n+2})+(-1)^{n+1}\mu\frac{u_{n}g_{n}(u_{n+1}g_{n+2}+u_{n+2}g_{n+1})}{h_{n-1}},\nonumber\\
&&b^{(-2)}=(-1)^{n}\mu c_{n-1}h_{n}h_{n-1}(u_{n+1}g_{n+2}+u_{n+2}g_{n+1}).\nonumber
\end{eqnarray*}
Here $\mu$ is an arbitrary constant, the factors $s_n^{(1)}$, $s_n^{(2)}$ are arbitrary periodic functions with the period 2. By substituting the coefficients into (\ref{nonaut_eq_L2}) one finds the searched operator $L_2$. Thus we have both operators $L_1$, $L_2$. We can conclude that the recursion operator has the form
\begin{eqnarray}\label{nonaut_eq_R}
&& R=r^{(2)}D_n^2+r^{(1)}D_n+r^{(0)}+r^{(-1)}D_n^{-1}+r^{(-2)}D_n^{-2}\nonumber\\
&& \qquad {} \qquad {} \quad {} +u_{n,\tau_1}(D_n-1)^{-1}\tilde{q}+u_{n,\tau_2}(D_n-1)^{-1}\tilde{p}
\end{eqnarray}
with undetermined factors $r^{(i)}$, $i=\overline{2,-2}$ and $\tilde{p}$, $\tilde{q}$. In order to find these factors we solve the equation $L_2=L_1R$ or the same
\begin{eqnarray}\label{nonaut_eq_searchR}
&& b^{(4)}D_n^4+b^{(3)}D_n^3+b^{(2)}D_n^2+b^{(1)}D_n+b^{(0)}+b^{(-1)}D_n^{-1}+b^{(-2)}D_n^{-2}=\nonumber\\
&& \qquad (\alpha D_n^2+\beta D_n+\gamma)\left(r^{(2)}D_n^2+r^{(1)}D_n+r^{(0)}+r^{(-1)}D_n^{-1}\right. \\
&& \qquad {} \quad {} \left.+r^{(-2)}D_n^{-2}+u_{n,\tau_1}(D_n-1)^{-1}\tilde{q}+u_{n,\tau_2}(D_n-1)^{-1}\tilde{p}\right). \nonumber
\end{eqnarray}
Comparing the coefficients of the same powers of the operator $D_n$, we find:
\begin{eqnarray*}\label{nonaut_eq_coeffR}
&&r^{(2)}=\mu c_nh_nh_{n-1}, \qquad r^{(1)}=\frac{\mu u_ng_n}{h_n},\nonumber\\
&&r^{(0)}=\frac{\mu(u_{n-1}g_n-u_ng_{n-1})}{h_{n-1}}+\mu c_n(h_{n-1}h_{n+1}-1)+\mu c_{n-1}(h_{n}h_{n-2}-1)\nonumber\\
&& \qquad {} \qquad {}-((-1)^ns_n^{(1)}+s_n^{(2)}),\nonumber\\
&&r^{(-1)}=-\frac{\mu u_ng_n}{h_{n-1}}, \qquad r^{(-2)}=\mu c_{n-1}h_nh_{n-1} \nonumber\\
&& \tilde{p}=\mu\left(\frac{u_{n-1}}{h_{n-1}}+\frac{u_{n+1}}{h_{n}}\right), \qquad \tilde{q}=(-1)^{n+1}\mu\left(\frac{g_{n-1}}{h_{n-1}}+\frac{g_{n+1}}{h_{n}}\right).
\end{eqnarray*}
Finally by taking $\mu=1$, we find the recursion operator 
\begin{eqnarray}\label{nonaut_eq_Rfinal}
&& R=c_nh_nh_{n-1}D_n^2+\frac{ u_ng_n}{h_n}D_n+\frac{u_{n-1}g_n-u_ng_{n-1}}{h_{n-1}}+ c_nh_{n-1}h_{n+1}+\nonumber\\
&& \quad  c_{n-1}h_{n}h_{n-2}+s_n-\frac{u_ng_n}{h_{n-1}}D_n^{-1}+ c_{n-1}h_nh_{n-1}D_n^{-2}+\label{Last}\\
&& \quad u_{n,\tau_1}(D_n-1)^{-1}(-1)^{n+1}\left(\frac{g_{n-1}}{h_{n-1}}+\frac{g_{n+1}}{h_{n}}\right)+u_{n,\tau_2}(D_n-1)^{-1}\left(\frac{u_{n-1}}{h_{n-1}}+\frac{u_{n+1}}{h_{n}}\right).\nonumber
\end{eqnarray}
where  $g_n=h_nh_{n-1}(c_nu_{n+2}-c_{n-1}u_{n-2})$, the factors $u_{n,\tau_1}$ and  $u_{n,\tau_2}$ are defined by (\ref{nonaut_eq_sym_1}) and, respectively, by (\ref{nonaut_eq_sym_2}). We denoted 
$s_n=(-1)^{n+1}s_n^{(1)}-s_n^{(2)}$. Obviously in the formula for $R$ function $s_n$ is considered as an arbitrary function satisfying the periodicity condition $s_n=s_{n+2}$. Recursion operator found in \cite{GYM14} is reduced to 
(\ref{Last}) with $s_n=0$.

\section*{Conclusions}  

The notion of a recursion operator is an important component of integrability. There are several methods for finding the recursion operator for a given integrable equation. In fact, they are all based either on the Lax representation, or on the multi-Hamiltonian approach. Here we proposed an alternative method based solely on the concept of symmetry. As a rule, the recursion operator is used to construct symmetries and conserved densities. We noticed that the recursion operator can be effectively obtained from a set of several first symmetries, which is called the set of seed symmetries. We look for the recursion operator as the ratio of two differential operators $ R = L_1 ^ {-1} L_2 $. Using the operators $L_1$ and $L_2$, one can easily define the Lax pair for the integrable nonlinear equation under consideration (see Formulas (\ref{psi-x}), (\ref{psi-t})).

\section*{Acknowledgments}

The authors gratefully acknowledge financial support from a Russian Science Foundation grant (project No 15-11-20007).

\end{document}